\documentclass[journal,10pt]{IEEEtran}
\IEEEoverridecommandlockouts 
\usepackage{amsmath,amssymb,amsfonts,mathtools,dsfont,multirow,relsize}
\usepackage{cite}
\usepackage{color}
\usepackage{caption,subcaption}
\usepackage{float}
\usepackage{hyperref}
\usepackage{algorithmicx}
\usepackage[ruled]{algorithm}
\usepackage{algpseudocode}
\usepackage{overpic}
\usepackage{amsmath, amsthm}

\newtheorem{lemma}{Lemma}

\newcommand{\bE}{\boldsymbol{E}}

\newcommand{\bH}{{\bf{H}}}

\newcommand{\bM}{\boldsymbol{M}}

\newcommand{\bS}{\boldsymbol{S}}

\newcommand{\bU}{\boldsymbol{U}}

\newcommand{\bV}{\boldsymbol{V}}

\newcommand{\bZ}{\boldsymbol{Z}}

\newcommand{\diag}{\mathrm{diag}}

\newcommand{\btheta}{\boldsymbol{\theta}}

\newcommand{\bomega}{\boldsymbol{\Omega}}

\newcommand{\bPsi}{\boldsymbol{\Psi}}

\newcommand{\bD}{\boldsymbol{\Delta}}

\newcommand{\Eta}{{\boldsymbol{\eta}}}

\newcommand{\bthe}{{\boldsymbol{\theta}}}

\begin{document}
\renewcommand{\thepage}{}

\title{Beyond Diagonal RIS Design for Parameter Estimation With and Without Eavesdropping}

\author{Özlem Tuğfe Demir and Sinan Gezici\thanks{Ö. T. Demir is with the Department of Electrical and Electronics
Engineering, TOBB University of Economics and Technology, Ankara,
Türkiye (Email: ozlemtugfedemir@etu.edu.tr). S. Gezici is with the Department of Electrical and Electronics Engineering, Bilkent University, Ankara 06800, Türkiye (Email: gezici@ee.bilkent.edu.tr).}\vspace{-0.8cm}}

\maketitle

\begin{abstract}
In this letter, we investigate the transmission of a complex-valued parameter vector from a transmitter to an intended receiver, considering both the presence and absence of an eavesdropper. The direct links from the transmitter to both the intended receiver and the eavesdropper are assumed to be blocked, and communications occur solely through cascaded channels facilitated by a beyond-diagonal reconfigurable intelligent surface (BD-RIS). While previous research has considered this system under conventional (diagonal) RIS assistance, we extend the setup to incorporate BD-RIS and quantify the resulting improvement in estimation performance at the intended receiver. This performance is measured by the trace of the Fisher information matrix (FIM), or equivalently, the average Fisher information, while simultaneously limiting the estimation capability of the eavesdropper. We propose solutions and algorithms for optimizing the BD-RIS response matrix and demonstrate their effectiveness. Numerical results reveal that the BD-RIS provides a significant enhancement in estimation quality compared to conventional diagonal RIS architectures.

\textbf{\textit{Index Terms}-- Beyond diagonal RIS (BD-RIS), Fisher information, parameter estimation, reconfigurable intelligent surface (RIS), Cram\'{e}r-Rao lower bound.}
\end{abstract}

\vspace{-2mm}
\section{Introduction}\label{sec:Intro}
\vspace{-1mm}

Eavesdropping stands as a significant security threat in wireless networks, primarily due to the inherently broadcast nature of the wireless medium \cite{he2022proactive,Abadi2024}. In fifth-generation (5G) wireless systems, delivering high-quality and secure communication services is a critical objective \cite{wu2019safeguarding}. Physical layer security (PLS) is a promising approach to ensure confidentiality, which leverages inherent randomness of wireless channels. Unlike traditional cryptographic methods, PLS facilitates secure communication without relying on encryption keys, operating directly at the physical transmission level \cite{wu2019safeguarding}. The primary objective of PLS-based approaches is to ensure secure communications with the intended receivers in the presence of eavesdroppers by exploiting the unique characteristics of wireless channels \cite{Abadi2024,yener2015wireless}.

Over the past decade, reconfigurable intelligent surfaces (RISs) have emerged as a key wireless technology for shaping the propagation environment to enhance critical performance metrics. Owing to their passive nature, RISs can manipulate the wireless channel by applying phase shifts to incident signals in an energy-efficient manner \cite{liu2021reconfigurable}. RIS has also been extensively explored in the context of PLS \cite{Abadi2024,khoshafa2024ris}. One of the recent studies has shown that optimized RIS phase profiles can enhance secure parameter estimation by maximizing the trace of the Fisher information matrix (FIM)—also known as the average Fisher information—at the legitimate receiver, while simultaneously limiting the average Fisher information gained by a potential eavesdropper \cite{Abadi2024}.  

Conventional RIS architectures employ single-connected elements with independent phase shifts, modeled by a diagonal reflection matrix where each diagonal entry modifies the impinging signal \cite{10316535}. Recent advances have introduced beyond-diagonal RIS (BD-RIS) architectures, where passive interconnections between elements enable non-diagonal reflection matrices. Shen et al. \cite{shen2021modeling} classify these as fully- or group-connected designs, offering greater flexibility and potential power gains under certain channel conditions \cite{shen2021modeling,zhou2023optimizing,demir2024wideband}.

Although RIS-assisted parameter estimation in the presence of eavesdropping has been studied in prior work \cite{Abadi2024}, the use of BD-RIS in this context remains unexplored. Unlike conventional diagonal RIS configurations with unit-modulus entries, BD-RIS allows for arbitrary unitary response matrices, with or without symmetry constraints—corresponding to reciprocal or non-reciprocal circuit designs, respectively. This added flexibility not only enhances design capabilities but also calls for new mathematical analyses (due to significantly more challenging and distinct optimization problems than those in \cite{Abadi2024}). In this letter, we address this gap by studying the parameter estimation problem for the first time under BD-RIS-assisted transmission. We propose several optimization solutions, either in closed-form or through efficient algorithms, for scenarios with and without eavesdropping. The resulting designs demonstrate substantial improvements in estimation quality at the intended receiver while effectively limiting the information gained by the eavesdropper.

\vspace{-2mm}

\section{System Model}
\vspace{-1mm}

We adopt the system model proposed in \cite{Abadi2024}, with the key difference that we consider assistance from a BD-RIS instead of a conventional diagonal RIS. A deterministic, $k$-dimensional complex-valued parameter vector, denoted by $\btheta = [\theta_1, \ldots, \theta_k]^T$, is transmitted from Alice (the transmitter) to Bob (the intended receiver). We analyze two distinct scenarios: (i) presence of an eavesdropper (Eve), and (ii) absence of an eavesdropper. The vector $\btheta$ is assumed to be unknown to both Bob and Eve. We assume that the direct links between Alice and both Bob and Eve are blocked, and Alice can reach Bob and Eve through the cascaded channels facilitated by the BD-RIS.
In this setting, measurements at Bob and Eve are obtained according to the following linear models \cite{Abadi2024}:
\begin{align} \label{y_bob}
    {\bf{y}}_b&=\left( {\bf{H}}_{rb}{\bf{\Omega}}{\bf{H}}_{ar} \right){\bf{P}}{\bthe}+\Eta_b\\ \label{y_eve}
    {\bf{y}}_e&=\left( {\bf{H}}_{re}{\bf{\Omega}}{\bf{H}}_{ar} \right){\bf{P}}{\bthe}+\Eta_e
\end{align}
where ${\bf{y}}_b \in \mathbb{C}^{n_b}$ and  ${\bf{y}}_e \in \mathbb{C}^{n_e}$ represent the received measurement vectors at Bob and Eve, respectively. The matrices 
${\bf{H}}_{ar} \in \mathbb{C}^{r\times k}$, ${\bf{H}}_{rb}\in \mathbb{C}^{n_b\times r}$, and ${\bf{H}}_{re}\in \mathbb{C}^{n_e\times r}$ denote the channel matrices corresponding to the Alice–RIS, RIS–Bob, and RIS–Eve links, respectively. The integers $k$, $n_b$, $n_e$, and $r$ denote the dimensionality of the parameter vector, the number of antenna elements at Bob, the number of antenna elements at Eve, and the number of RIS elements, respectively. The noise vectors $\Eta_b \in \mathbb{C}^{n_b}$ and $\Eta_e \in \mathbb{C}^{n_e}$ are modeled as circularly symmetric complex Gaussian with distributions 
${\Eta}_b \sim \mathcal{CN}(\boldsymbol{0},\boldsymbol{\Sigma}_{b})$ and 
${\Eta}_e \sim \mathcal{CN}(\boldsymbol{0},\boldsymbol{\Sigma}_{e})$, where
$\boldsymbol{\Sigma}_{b}$ and
$\boldsymbol{\Sigma}_{e}$ are positive definite covariance matrices. The power allocation matrix is given by ${\bf{P}} = \diag \{ \sqrt{p_1},\sqrt{p_2}, \dots,\sqrt{p_k}  \}$ and is assumed to be fixed. Finally, ${\bf{\Omega}}$ is the $r\times r$ BD-RIS response matrix.

As in \cite{Abadi2024}, the trace of the FIM, which indicates the overall usefulness of a measurement for vector parameter estimation \cite{Gurgunoglu}, is considered as the estimation performance metric due to its tractable and intuitive form, which can be calculated for Bob and Eve as \cite[Eq.~(14)]{Abadi2024}
\begin{align}\label{eq:AvgFIbob}
{\rm{tr}}\left(  {\bf{I}}({\bf{y}}_b;{\bthe}) \right)
&={\rm{tr}} \left(({\bf{H}}_{rb}{\bf{\Omega}}{\bf{H}}_{ar}{\bf{P}} )^H
        \boldsymbol{\Sigma}_b^{-1}({\bf{H}}_{rb}{\bf{\Omega}}{\bf{H}}_{ar}{\bf{P}} ) \right)  
        \\\label{eq:AvgFIeve}
{\rm{tr}}\left(  {\bf{I}}({\bf{y}}_e;{\bthe}) \right)
&= {\rm{tr}} \left(({\bf{H}}_{re}{\bf{\Omega}}{\bf{H}}_{ar}{\bf{P}} )^H
        \boldsymbol{\Sigma}_e^{-1}({\bf{H}}_{re}{\bf{\Omega}}{\bf{H}}_{ar}{\bf{P}} ) \right)
\end{align}

\section{BD-RIS Design without Eavesdropping}

To begin with, we omit the eavesdropper and only consider the intended user, Bob. The trace of the FIM for Bob can be written from \eqref{eq:AvgFIbob} as
\begin{align}
{\rm{tr}}\left(  {\bf{I}}({\bf{y}}_b;{\bthe}) \right)
&={\rm{tr}} \left({\bf{P}}^H{\bf{H}}_{ar}^H{\bf{\Omega}}^H{\bf{H}}_{rb}^H
\boldsymbol{\Sigma}_b^{-1}{\bf{H}}_{rb}{\bf{\Omega}}{\bf{H}}_{ar}{\bf{P}} \right)  
\\
&={\rm{tr}} \left(\bH^H{\bf{\Omega}}^H\bE_b{\bf{\Omega}}\bH \right)  
\end{align}
with $\bE_b\triangleq{\bf{H}}_{rb}^H
\boldsymbol{\Sigma}_b^{-1}{\bf{H}}_{rb}\in{\mathbb{C}}^{r\times r}$ and $\bH\triangleq {\bf{H}}_{ar}{\bf{P}}\in{\mathbb{C}}^{r\times k}$. We first focus on the following optimization problem: 
\begin{subequations}\label{eq:OptProbGenel1}
\begin{align}
&\underset{\bomega}\max~~~~~~~~~{\rm{tr}}\left(\bH^H\bomega^H\bE_b\bomega\bH\right)
    \\
    &{\rm{subject~to~}}~\bomega^H\bomega={\bf{I}}_{r}
\end{align}
\end{subequations}
where ${\bf{I}}_{r}$ is the $r\times r$ identity matrix and $\bomega$ is the $r\times r$ BD-RIS response matrix, specified as a complex unitary matrix in the constraint. Here, we consider the most general form of a fully connected BD-RIS, where the reciprocity constraint (symmetric $\bomega$) is relaxed. In fact, non-reciprocal circuit networks can practically be realized, and solving this problem yields the ultimate limit on the average Fisher information—analogous to capacity limits in communication systems \cite{bjornson2024capacity}. 

We can express the optimization problem in \eqref{eq:OptProbGenel1} as
\begin{subequations}\label{eq:optimization1}
\begin{align}\label{eq:ObjFn2}
&\underset{\bomega}\max~~~~~~~~~{\rm{tr}}\left(\bomega^H\bE_b\bomega\bM\right)
    \\\label{eq:Const2}
    &{\rm{subject~to~}}~\bomega^H\bomega={\bf{I}}_{r}
\end{align} 
\end{subequations}
where $\bM\triangleq\bH\bH^H\in{\mathbb{C}}^{r\times r}$. The solution of \eqref{eq:optimization1} is specified in the following lemma.

\begin{lemma} The optimal BD-RIS response matrix for the problem \eqref{eq:optimization1} is given as 
\begin{align}
    \bomega = \bV_E \bV_M^H
\end{align}
where $\bV_E$ and $\bV_M$ are the unitary matrices with the columns as the eigenvectors of $\bE_b$ and $\bM$, respectively. The eigenvectors are arranged to correspond to the eigenvalues, which are sorted in descending order.

\begin{proof}
    We use $\bE_b=\bV_E\bD_E\bV_E^H$ and $\bM= \bV_M \bD_M\bV_M^H$ for denoting the eigenvalue decompositions of $\bE_b$ and $\bM$, respectively. The diagonal matrices $\bD_E$ and $\bD_M$ have the eigenvalues arranged in descending order along their diagonals. Since $\bomega^H\bE_b\bomega$ and $\bM$ are positive semi-definite matrices and the eigenvalues of $\bomega^H\bE_b\bomega$ and $\bE_b$ are the same (due to unitary $\bomega$), we have from Von Neumann's trace inequality the following:
    \begin{align}        {\rm{tr}}\left(\bomega^H\bE_b\bomega\bM\right)\leq \sum_{i=1}^r \delta_{E,i}\delta_{M,i}
    \end{align}
    where $\delta_{E,1}\geq \delta_{E,2} \geq \cdots \geq \delta_{E,r}$ and $\delta_{M,1}\geq \delta_{M,2}\geq \cdots \geq \delta_{M,r}$ are the sorted eigenvalues in $\bD$ and $\boldsymbol{\Sigma}$, respectively. We then observe that inserting $\bomega = \bV_E \bV_M^H$ results in this upper bound. Hence, it is an optimal configuration.
\end{proof}
\end{lemma}

Now, we consider the average Fisher information maximization problem for a reciprocal BD-RIS by adding the symmetry constraint as 
\begin{subequations}
\begin{align}\label{eq:ObjFn2_2}
&\underset{\bomega}\max~~~~~~~~~{\rm{tr}}\left(\bomega^H\bE_b\bomega\bM\right)
    \\\label{eq:Const2_2}
    &{\rm{subject~to~}}~\bomega^H\bomega={\bf{I}}_{r}
    ~{\rm{and}}~\bomega=\bomega^T
\end{align} \label{eq:optimization2}
\end{subequations}

This problem can be solved using manifold optimization techniques as in \cite{santamaria2024mimo}, where the overall steps of the main alternating optimization (AO) algorithm are outlined in Algorithm~\ref{alg1}. The initialization point is taken as $\bomega_0 = \bU\bU^{T}$, where $\bU$ is the unitary matrix obtained from the Takagi singular value decomposition of $\bV_E\bV_M^{H}+\bV_M^*\bV_E^{T}$
as
\begin{align}
\bV_E\bV_M^{H}+\bV_M^*\bV_E^{T} = \bU \boldsymbol{\Sigma}\bU^{T}. \label{eq:takagi}
\end{align}
As derived in \cite[The.~1]{demir2024wideband}, $\bomega_0 =\bU\bU^{T}$ is the closest entry of the set determined by $\bomega^H\bomega={\bf{I}}_{r}
    ~{\rm{and}}~\bomega=\bomega^T$ to $\bV_E\bV_M^{H}$ which is the optimal configuration when there is no reciprocity constraint.

\begin{algorithm}[h!] 
	\caption{AO Algorithm for Solving \eqref{eq:optimization2}.} \label{alg1}
	\begin{algorithmic}[1]
		\State {\bf Initialization:} Select $\bomega_{0}=\bU\bU^{T}$, where $\bU$ is obtained from \eqref{eq:takagi}. Set the iteration counter to zero, i.e., $i=0$. Convergence threshold: $\epsilon$. Learning step size $\mu>0$
        \While{Improvement in the cost function larger than $\epsilon$}
        \State Compute the gradient of the cost function with respect to $\bU^*$ as
        \begin{align}
    \bZ= \bE_b\bU_{i}\bU_{i}^{T}\bM\bU_{i}^*    
        \end{align}
        \State Compute $\bU_{i}e^{\mu\bS_{\rm skew}}$, where $e^{\mu\bS_{\rm skew}}$ is the matrix exponential and
        \begin{align}
        \bS_{\rm skew}=\frac{\bU_i^{H}\bZ-\bZ^{H}\bU_i}{2}\cdot
        \end{align}
        \State If there is a cost improvement, increase $\mu$, update $\bU_{i+1}$ as $\bU_{i}e^{\mu\bS_{\rm skew}}$ and cost function. Otherwise, decrease $\mu$ and do not change $\bU_{i}$ and the cost.
        \State $i\leftarrow i+1$
        \EndWhile
		\State {\bf Output:} $\bomega=\bU_{i}\bU_{i}^{T}$
	\end{algorithmic}
\end{algorithm}

\section{Secure BD-RIS Design in the Presence of Eavesdropping}

In this section, there also exists a eavesdropper, Eve, in the environment. The average Fisher information at Eve can be expressed via \eqref{eq:AvgFIeve} as
\begin{align}
{\rm{tr}}\left(  {\bf{I}}({\bf{y}}_e;{\bthe}) \right)
&={\rm{tr}} \left({\bf{P}}^H{\bf{H}}_{ar}^H{\bf{\Omega}}^H{\bf{H}}_{re}^H
\boldsymbol{\Sigma}_e^{-1}{\bf{H}}_{re}{\bf{\Omega}}{\bf{H}}_{ar}{\bf{P}} \right) \nonumber 
\\
&{\hspace{-0.15cm}}={\rm{tr}} \left(\bH^H{\bf{\Omega}}^H\bE_e{\bf{\Omega}}\bH \right) = {\rm{tr}}\left(\bomega^H\bE_e\bomega\bM\right) 
\end{align}
where $\bE_e\triangleq{\bf{H}}_{re}^H
\boldsymbol{\Sigma}_e^{-1}{\bf{H}}_{re}$. Then, the optimization problem in the presence of eavesdropping for non-reciprocal BD-RIS (the most general form) can be written as
\begin{subequations}\label{eq:optMostGen}
\begin{align}\label{eq:ObjFn-eave}
&\underset{\bomega}\max~~~~~~~~~{\rm{tr}}\left(\bomega^H\bE_b\bomega\bM\right)
    \\\label{eq:Const2-eave}
    &{\rm{subject~to~}}~\bomega^H\bomega={\bf{I}}_{r} \\
    & \hspace{18mm} {\rm{tr}}\left(\bomega^H\bE_e\bomega\bM\right)\leq \epsilon
\end{align}
\end{subequations}
where we limit the average Fisher information at Eve by $\epsilon>0$. 

We first re-express the problem in \eqref{eq:optMostGen} as
\begin{subequations}\label{eq:optMostGen2}
\begin{align}\label{eq:ObjFn-eave2}
&\underset{\bomega,\bPsi}\max~~~~~~~~~\Re\left({\rm{tr}}\left(\bomega^H\bE_b\bPsi\bM\right)\right)
    \\\label{eq:Const2-eave2}
    &{\rm{subject~to~}}~\bomega^H\bomega={\bf{I}}_{r} \\
    &\hspace{18mm} \bPsi=\bomega \\
    & \hspace{18mm} {\rm{tr}}\left(\bPsi^H\bE_e\bPsi\bM\right)\leq \epsilon.
\end{align}
\end{subequations}
Then, we adopt the penalty dual decomposition (PDD) method \cite{Shi2020,zhou2023optimizing} to solve the problem in \eqref{eq:optMostGen2}. To this end, the augmented Lagrangian function is obtained as
\begin{align}
   \mathcal{L}(\bomega,{\bf \Psi},{\bf \Lambda})&=-\Re\left({\rm{tr}}\left(\bomega^H\bE_b\bPsi\bM\right)\right)+ \frac{1}{2\rho}\Vert \bomega-{\bf \Psi}\Vert_F^2 \nonumber\\
   &\quad + \Re\left({\rm{tr}}\left({\bf \Lambda}^H(\bomega-{\bf \Psi})\right)\right)
\end{align}
where ${\bf \Lambda}$ is the Lagrangian dual variable associated with the constraint ${\bf \Psi}={\bf \Omega}$ and $\rho^{-1}$ is the penalty coefficient. The steps outlined in the following subsections are repeated until convergence as described in \cite[Alg.~1]{zhou2023optimizing} with the major difference in the inner layer updates as outlined below.

\subsection{Update ${\bf \Omega}$ of  PDD Inner Layer Iteration}

The augmented Lagrangian minimization problem in terms of $\bomega$ can be written as
\begin{subequations}
\begin{align}\label{eq:ObjFn-eave3a}
&\underset{\bomega}\min~~~~~~~~~\left\Vert \bomega-\left({\bf \Psi}+\rho\bE_b\bPsi\bM-\rho{\bf \Lambda}\right)\right \Vert_F^2
    \\\label{eq:Const2-eave3a}
    &{\rm{subject~to~}}~\bomega^H\bomega={\bf{I}}_{r} 
\end{align}
\end{subequations}
which can analytically be solved using \cite[Lemma 1]{zhou2023optimizing}.

\subsection{ Update ${\bf \Psi}$ of  PDD Inner Layer Iteration}

The augmented Lagrangian minimization problem in terms of ${\bf \Psi}$ can be written as
\begin{subequations}
\begin{align}\label{eq:ObjFn-eave3b}
&\underset{\bPsi}\min~~~~~~~~~\left\Vert \bPsi-\left(\bomega+\rho\bE_b^H\bomega\bM^H+\rho{\bf \Lambda}\right)\right \Vert_F^2
    \\\label{eq:Const2-eave3b}
    &{\rm{subject~to~}}~{\rm{tr}}\left(\bPsi^H\bE_e\bPsi\bM\right)\leq \epsilon 
\end{align}
\end{subequations}
which can be cast as a quadratically-constrained quadratic programming form by vectorizing the matrices. Using the vec Kronecker identity in \cite{magnus2019matrix}, we can express the problem as
\begin{subequations}
\begin{align}\label{eq:ObjFn-eave4b}
&\underset{{\bf x}}\min~~~~~~~~~\left\Vert {\bf x}-{\bf b}\right \Vert^2
    \\\label{eq:Const2-eave4b}
    &{\rm{subject~to~}}~{\bf x}^H{\bf A}{\bf x}\leq \epsilon 
\end{align}
\end{subequations}
where ${\bf b}= \mathrm{vec}\left(\bomega+\rho\bE_b^H\bomega\bM^H+\rho{\bf \Lambda}\right)$, ${\bf x}=\mathrm{vec}(\bPsi)$, and ${\bf A}=\bM^T\otimes \bE_e$.

To solve the above problem in semi-closed form, we first express the eigenvalue decomposition of ${\bf A}$ as ${\bf A}=\sum_{i=1}^{r^2}\lambda_i{\bf u}_i{\bf u}_i^H$, where $\lambda_i\geq 0$ is the $i$th eigenvalue with the corresponding unit-norm eigenvector ${\bf u}_i\in \mathbb{C}^{r^2}$. We then express the optimization variable as ${\bf x}=\sum_{i=1}^{r^2}\alpha_ie^{j\phi_i}{\bf u}_i$, where $\alpha_i>0$ and the phase-shift $\phi_i$ are the newly defined optimization variables. The problem can be then written as
\begin{subequations}
\begin{align}\label{eq:ObjFn-eave5b}
&\underset{\{\alpha_i,\phi_i\}}\min~~~~~~~~~\left\Vert \sum_{i=1}^{r^2}\alpha_ie^{j\phi_i}{\bf u}_i-{\bf b}\right \Vert^2
    \\\label{eq:Const2-eave5b}
    &{\rm{subject~to~}}~\sum_{i=1}^{r^2}\lambda_i\alpha_i^2\leq \epsilon 
\end{align}
\end{subequations}
which is equivalent to
\begin{subequations}\label{eq:OptPhaseProb}
\begin{align}\label{eq:ObjFn-eave6b}
&\underset{\{\alpha_i,\phi_i\}}\min~~~~~~~~~\sum_{i=1}^{r^2}\alpha_i^2-\sum_{i=1}^{r^2}2\alpha_i\Re\left(e^{\phi_i}{\bf b}^H{\bf u}_i\right)
    \\\label{eq:Const2-eave6b}
    &{\rm{subject~to~}}~\sum_{i=1}^{r^2}\lambda_i\alpha_i^2\leq \epsilon 
\end{align}
\end{subequations}
In \eqref{eq:OptPhaseProb}, the optimal phase-shifts are obtained as
\begin{align}
    \phi_i = \angle{{\bf u}_i^H{\bf b}}
\end{align}
which leads to the optimization problem only in terms of $\{\alpha_i\}$ as
\begin{subequations}\label{eq:OptAlphas}
\begin{align}\label{eq:ObjFn-eave7b}
&\underset{\{\alpha_i\}}\min~~~~~~~~~\sum_{i=1}^{r^2}\alpha_i^2-\sum_{i=1}^{r^2}2\alpha_i\left|{\bf u}_i^H{\bf b}\right|
    \\\label{eq:Const2-eave7b}
    &{\rm{subject~to~}}~\sum_{i=1}^{r^2}\lambda_i\alpha_i^2\leq \epsilon 
\end{align}
\end{subequations}
The Lagrangian function for \eqref{eq:OptAlphas} is given by
\begin{equation}
    \mathcal{L}(\{\alpha_i\}, \mu) = \sum_{i=1}^{r^2} \alpha_i^2 - \sum_{i=1}^{r^2} 2\alpha_i \left|{\bf u}_i^H{\bf b}\right|
    + \mu \left( \sum_{i=1}^{r^2} \lambda_i\alpha_i^2 - \epsilon \right)
\end{equation}
where $\mu \geq 0$ is the Lagrange multiplier. Then, the Karush-Kuhn-Tucker (KKT) conditions are stated as follows:
   \begin{equation}
       \frac{\partial \mathcal{L}}{\partial \alpha_i} = 2\alpha_i - 2 \left|{\bf u}_i^H{\bf b}\right| + 2\mu \lambda_i \alpha_i = 0,
   \end{equation}
which yields 
   \begin{equation}
       \alpha_i = \frac{\left|{\bf u}_i^H{\bf b}\right|}{1 + \mu \lambda_i},
   \end{equation}
   \begin{equation}
       \sum_{i=1}^{r^2} \lambda_i\alpha_i^2 \leq \epsilon,
   \end{equation}
   \begin{equation}
       \mu \geq 0,
   \end{equation}
   \begin{equation}
       \mu \left( \sum_{i=1}^{r^2} \lambda_i\alpha_i^2 - \epsilon \right) = 0.
   \end{equation}

To determine the optimal value of $\mu$, we first set $\mu = 0$, then $\alpha_i = \left|{\bf u}_i^H{\bf b}\right|$, and we check if the constraint holds.
If $\sum_{i=1}^{r^2} \lambda_i \left|{\bf u}_i^H{\bf b}\right|^2 \leq \epsilon$, then this is the optimal solution. Otherwise, we solve for $\mu$ such that
  \begin{equation}
      \sum_{i=1}^{r^2} \lambda_i \left(\frac{\left|{\bf u}_i^H{\bf b}\right|}{1 + \mu \lambda_i}\right)^2 = \epsilon.
  \end{equation}
This equation in $\mu$ can be solved via a bisection search.

\subsection{Update of PDD Outer Layer Iteration}

For updating the PDD outer layer iteration, the procedure in \cite[Eqns.~(31)-(32)]{zhou2023optimizing} is applied.

When solving the optimization problem for the reciprocal BD-RIS, an additional symmetry constraint arises, which affects only the update of ${\bf \Omega}$ in the inner iteration of the PDD algorithm. This update is implemented by solving
\begin{subequations}
\begin{align}
&\underset{\bomega}\min~~~~~~~~~\left\Vert \bomega-\left({\bf \Psi}+\rho\bE_b\bPsi\bM-\rho{\bf \Lambda}\right)\right \Vert_F^2
    \\
    &{\rm{subject~to~}}~\bomega^H\bomega={\bf{I}}_{r}  ~{\rm{and}}~\bomega=\bomega^T 
\end{align}
\end{subequations}
which can analytically be solved using \cite[The.~1]{demir2024wideband}.

\section{Numerical Results}

In this section, we quantify the performance improvements achieved via BD-RIS over the conventional diagonal RIS by leveraging the proposed theoretical and algorithmic solutions to the considered optimization problems. Following the methodology in \cite{Abadi2024}, the real and imaginary parts of the channel matrices are generated as independent and identically distributed (i.i.d.) random variables uniformly drawn from the interval  $[-0.1,0.1]$, based on a single realization in MATLAB. The covariance matrices of the additive noise vectors at Bob and Eve are set to $10^{-5}$ times the identity matrix. The number of antennas at both Bob and Eve is set as $n_b=n_e=2k$, where $k$ is the number of parameters. The power allocation matrix is set to $\sqrt{30/k}$ times identity corresponding to equal power allocation. 

For diagonal RIS assisted design in the absence of eavesdropping, the semidefinite programming (SDP) problem without the constraint related to Eve, as formulated in \cite{Abadi2024}, is solved, and Gaussian randomization is applied to extract a rank-one solution. In the presence of eavesdropping, it is observed that small values of $\epsilon$—corresponding to a stringent estimation quality threshold for Eve—often result in infeasible problems when the RIS response matrix is constrained to have unit-modulus diagonal entries. To address this issue, we relax the constraint to allow diagonal entries with magnitudes less than or equal to one as in \cite[Sec.~V.B]{Abadi2024} and solve the corresponding positive semidefinite programming problem. After solving that problem, Gaussian randomization is performed, and any diagonal entry of the RIS response matrix exceeding unit magnitude is clipped to have unit modulus.

\begin{figure}[t!]
        \centering
	\begin{overpic}[width=.95\columnwidth,tics=10]{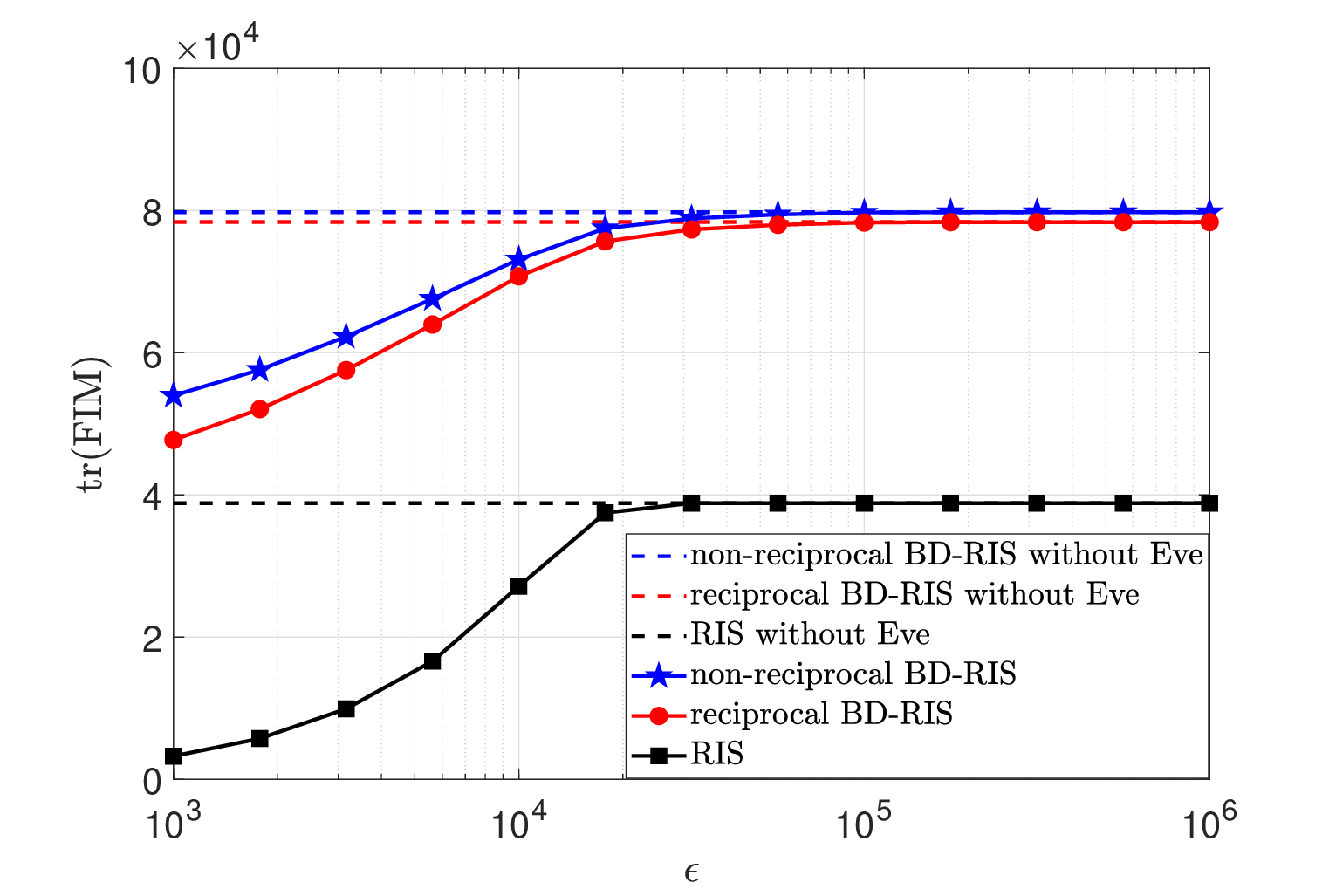}
\end{overpic} 
\vspace{-3mm}
        \caption{Trace of FIM at Bob achieved by non-reciprocal BD-RIS, reciprocal BD-RIS, and conventional RIS in both the presence and absence of eavesdropping for $r=36$ and $k=10$.}
        \label{fig:1}
        \vspace{-4mm}
\end{figure}

In Fig.~\ref{fig:1}, we set the number of BD-RIS and RIS elements to $r=36$, and the number of parameters to be estimated to $k=10$. We compare three configurations: i) non-reciprocal BD-RIS without symmetry constraints on the response matrix; ii) reciprocal BD-RIS with symmetry constraints; and iii) conventional diagonal RIS. The $x$-axis of the figure represents the average Fisher information limit of Eve, denoted by $\epsilon$. The dashed lines indicate the average Fisher information at Bob in the absence of eavesdropping; since they do not depend on $\epsilon$, they represent the ultimate achievable average Fisher information limit. The solid curves correspond to scenarios with eavesdropping. As $\epsilon$ increases, the average information at Bob approaches the corresponding ultimate limit. As expected, the highest information is obtained with the non-reciprocal BD-RIS, as it represents the most general form of BD-RIS. Imposing the reciprocity constraint leads to a slight reduction in average Fisher information, although the degradation is negligible. In contrast, when employing a conventional diagonal RIS, the Fisher information is significantly reduced, especially when Eve's information constraint is stringent (i.e., small $\epsilon$). These results demonstrate that BD-RIS notably enhances the estimation performance at Bob even under tight information leakage constraints to Eve.

\begin{figure}[t!]
        \centering
	\begin{overpic}[width=.95\columnwidth,tics=10]{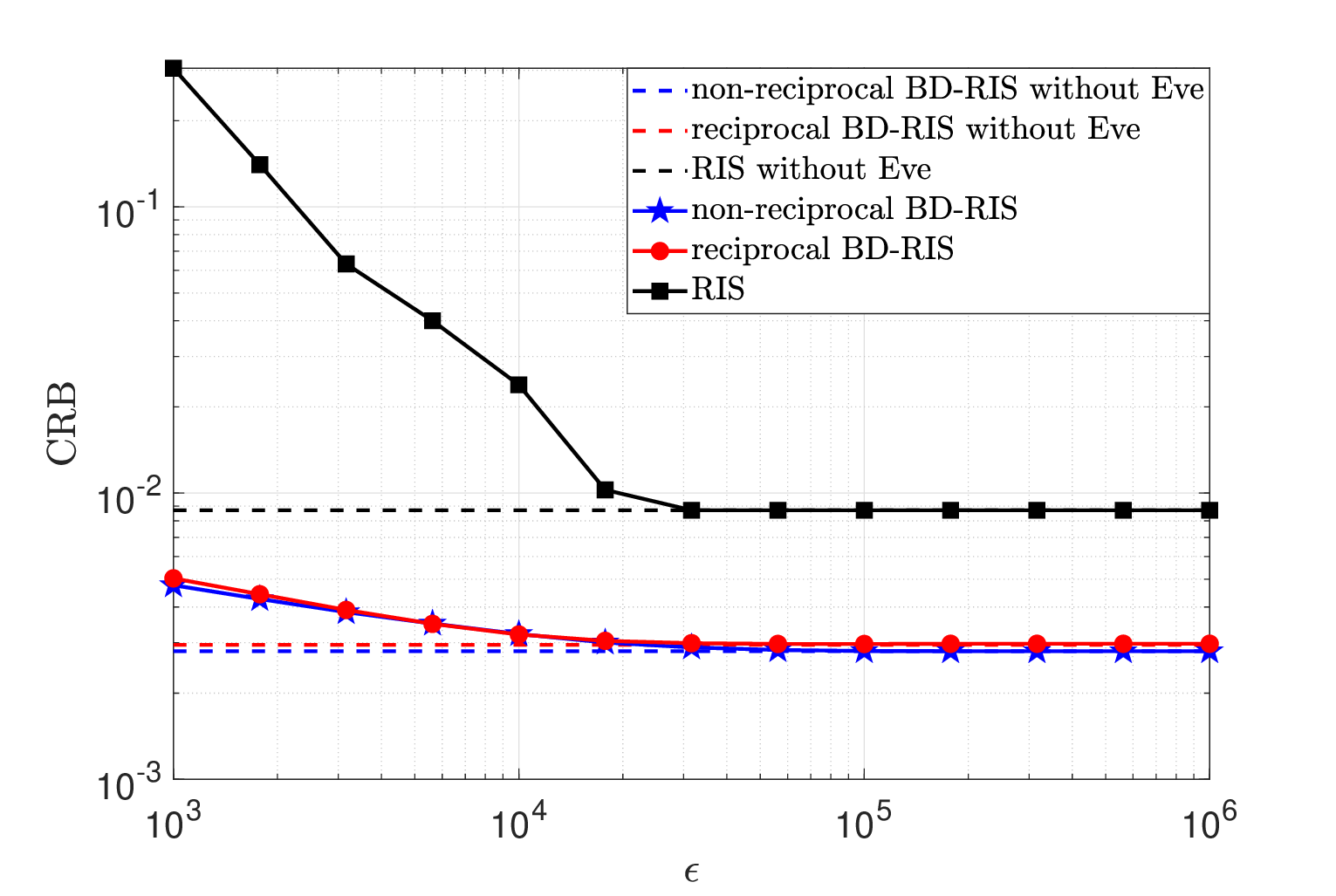}
\end{overpic} 
\vspace{-3mm}
        \caption{CRB (equivalently, MSE of the MLE) at Bob  achieved by non-reciprocal BD-RIS, reciprocal BD-RIS, and conventional RIS in both the presence and absence of eavesdropping for $r=36$ and $k=10$. }
        \label{fig:2}
        \vspace{-4mm}
\end{figure}

In Fig.~\ref{fig:2}, we plot the Cramér–Rao bound (CRB), defined as the trace of the inverse FIM using the same setup as in Fig.~\ref{fig:1}. Since the maximum likelihood estimator (MLE) achieves the CRB for linear models with additive Gaussian noise as in \eqref{y_bob} and \eqref{y_eve}, the $y$-axis of Fig.~\ref{fig:2} also corresponds to the mean-squared error (MSE) of the MLE at Bob. As the figure illustrates, the BD-RIS significantly reduces the CRB, indicating improved estimation accuracy in terms of MSE. This improvement is particularly pronounced when the information leakage to Eve is tightly constrained (i.e., when $\epsilon$ is small). Moreover, BD-RIS enables the system to maintain a favorable CRB even under stringent leakage constraints, demonstrating its robustness in secure parameter estimation.

\begin{figure}[t!]
        \centering
	\begin{overpic}[width=.95\columnwidth,tics=10]{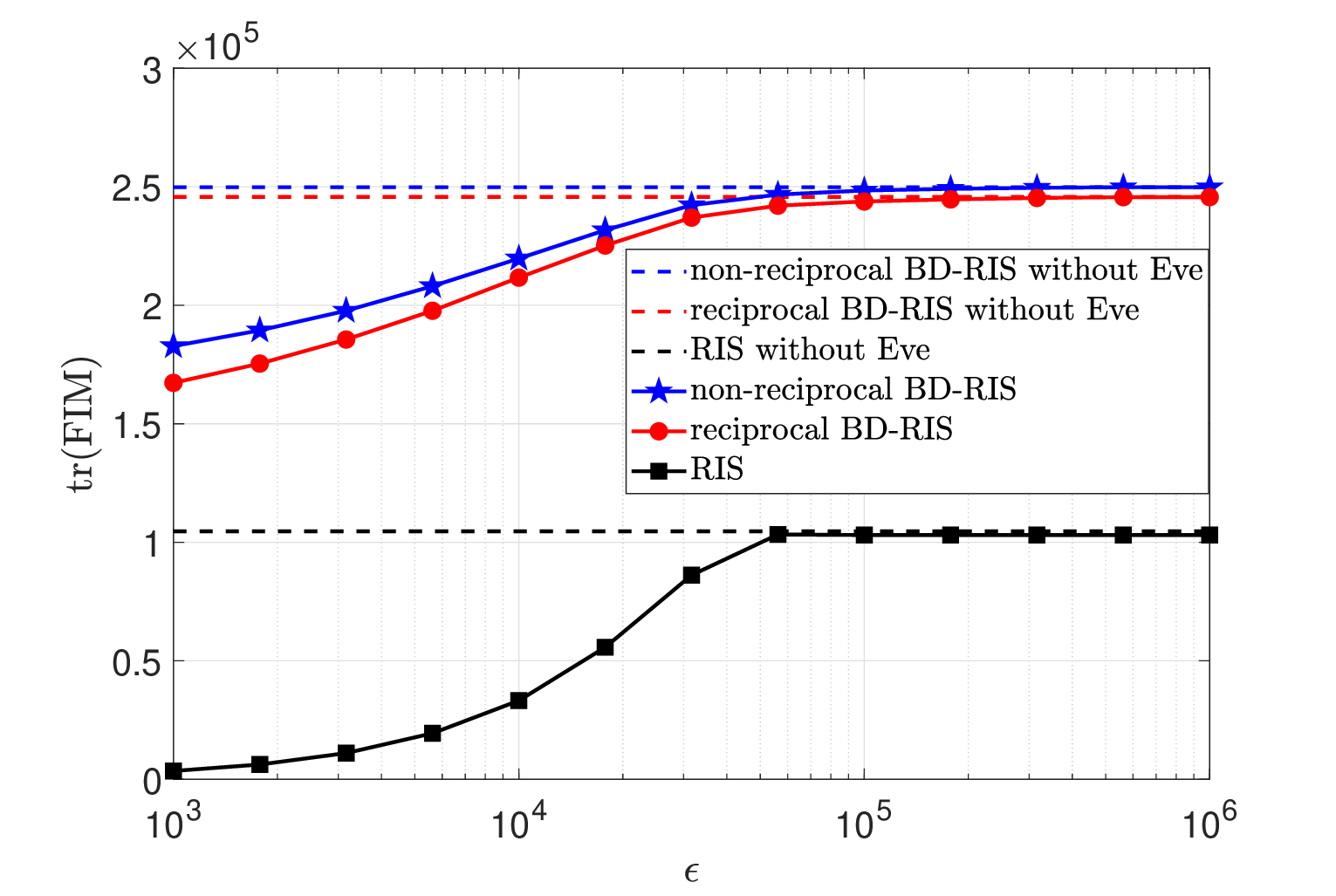}
\end{overpic} 
\vspace{-3mm}
        \caption{Trace of FIM at Bob  achieved by non-reciprocal BD-RIS, reciprocal BD-RIS, and conventional RIS in both the presence and absence of eavesdropping for $r=64$ and $k=15$.}
        \label{fig:3}
        \vspace{-4mm}
\end{figure}

\begin{figure}[t!]
        \centering
	\begin{overpic}[width=.95\columnwidth,tics=10]{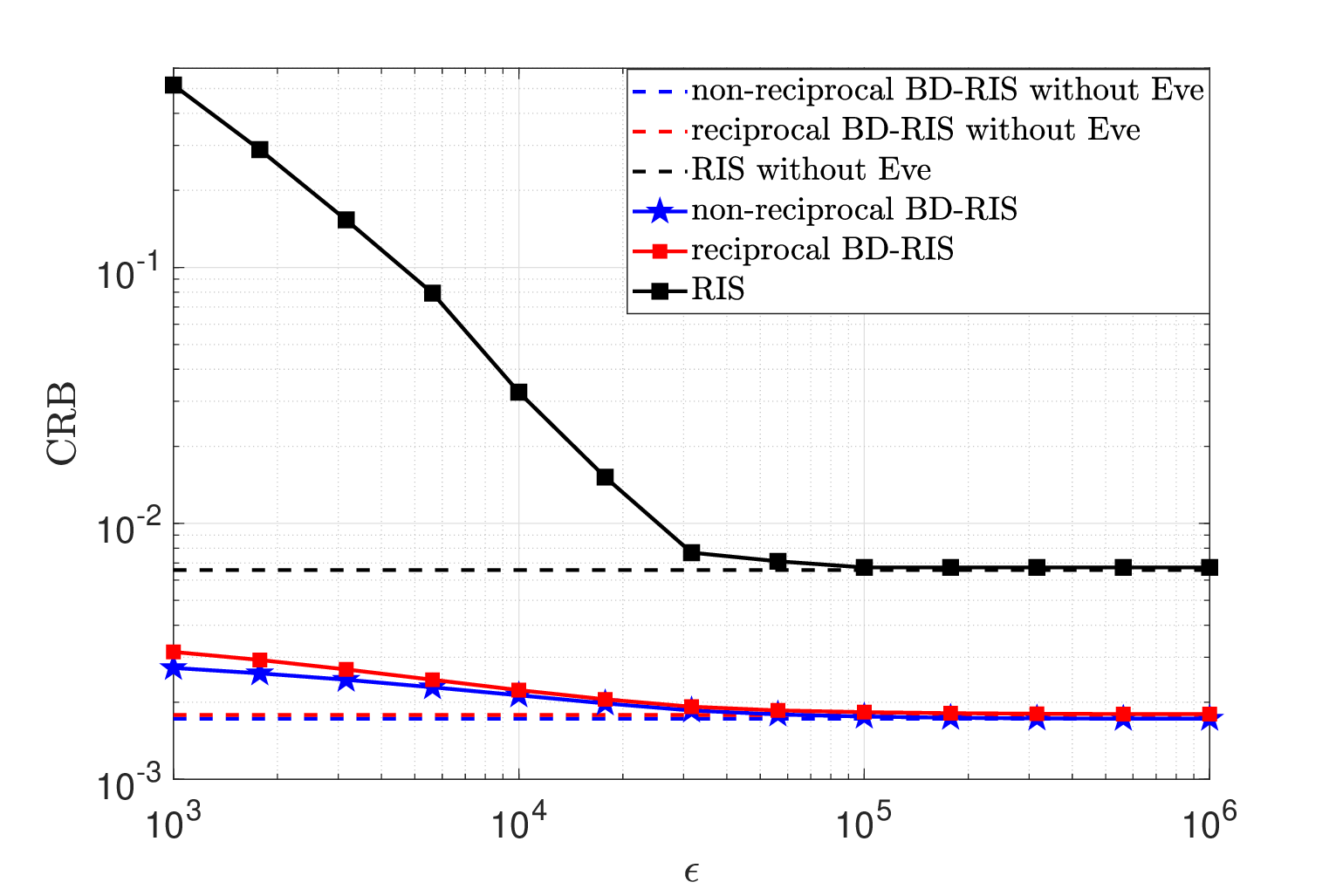}
\end{overpic} 
\vspace{-3mm}
        \caption{CRB (equivalently, MSE of the MLE) at Bob  achieved by non-reciprocal BD-RIS, reciprocal BD-RIS, and conventional RIS in both the presence and absence of eavesdropping for $r=64$ and $k=15$.}
        \label{fig:4}
        \vspace{-4mm}
\end{figure}

In Figs.~\ref{fig:3} and \ref{fig:4}, we repeat the previous experiments with a larger BD-RIS and RIS size ($r=64$) and an increased number of parameters ($k=15$). As shown in the figures, the estimation quality is further enhanced when using BD-RIS compared to conventional RIS. The performance gap between the non-reciprocal and reciprocal BD-RIS designs remains negligible, indicating that the more easily implementable reciprocal circuit networks offer a practical and effective alternative for BD-RIS deployment. Once again, we observe that BD-RIS enables robust and secure estimation even in the presence of eavesdropping.

\section{Conclusion}

In this letter, we proposed a BD-RIS architecture for secure parameter estimation in wireless systems, considering both the presence and absence of an eavesdropper. By formulating and solving optimization problems for the BD-RIS response matrix under different constraints—non-reciprocal, reciprocal, and diagonal—we quantified the fundamental performance limits using the trace of the FIM. Our results demonstrate that BD-RISs significantly outperform conventional diagonal RISs, particularly in scenarios with stringent eavesdropping constraints. Furthermore, the performance degradation due to the reciprocity constraint is shown to be negligible, making reciprocal BD-RIS a practical and effective alternative. These findings highlight the strong potential of BD-RIS in enhancing secure and accurate wireless parameter estimation.

\bibliographystyle{IEEEtran}
\bibliography{IEEEabrv,refs}

\end{document}